\documentclass[prl,aps,twocolumn,floatfix,showpacs]{revtex4}
\usepackage{graphicx,graphics,psfrag,amsmath,calc}
\usepackage{epsfig}
\usepackage{color}
\begin{document}

\title{
Formation of Feshbach molecules in the presence of \\
artificial spin-orbit coupling and Zeeman fields
}

\author{Doga Murat Kurkcuoglu and C. A. R. S\'a de Melo}

\affiliation{
School of Physics, Georgia Institute of Technology, 
Atlanta, 30332, USA
}

\date{\today}

\begin{abstract}
We derive general conditions for the emergence of singlet Feshbach molecules 
in the presence of artificial Zeeman fields for arbritary mixtures of Rashba 
and Dresselhaus spin-orbit orbit coupling in two or three dimensions. 
We focus on the formation of two-particle bound states resulting 
from interactions between ultra-cold spin-$1/2$ fermions, under 
the assumption that interactions are short-ranged 
and occur only in the s-wave channel. In this case, we calculate 
explicitly binding energies of Feshbach molecules and analyze 
their dependence on spin-orbit couplings, Zeeman fields,
interactions and center of mass momentum, 
paying particular attention to the experimentally relevant
case of spin-orbit couplings with equal Rashba and Dresselhaus (ERD) 
amplitudes. 
\end{abstract}


\maketitle

%
%% Introduction
%

The effects of spin-orbit interactions is ubiquitous in nature, from the 
macroscopic scale of the Earth-Moon complex in astronomy and astrophysics, 
to the microscopic scale of the electron in the hydrogen atom in atomic
physics. The interest in spin-orbit coupled systems has been revived 
in condensed matter physics due the emergence of non-trivial topological 
properties of insulators and superconductors subject to Rashba spin-orbit 
fields~\cite{kane-2010,s-zhang-2011}, 
and in atomic physics due to the creation of 
artificial spin-orbit coupling in ultra-cold atoms~\cite{spielman-2011},
which made possible the study of special quantum phase transitions in 
bosonic systems.

This new tool in the toolbox of atomic physics was experimentally 
developed first to study interacting bosonic atoms
where an equal Rashba-Dresselhaus (ERD) artificial spin-orbit 
coupling was created~\cite{spielman-2011}. It was suggested that 
interacting fermions could be studied using the 
same technique~\cite{spielman-2011, chapman-2011}. 
Estimulated by the dense literature of the effects of Rashba 
spin-orbit coupling (SOC) encountered in condensed 
matter physics~\cite{kane-2010,s-zhang-2011}, 
several theoretical groups investigated the effects of Rashba SOC 
for interacting ultra-cold fermions using mean field 
theories~\cite{shenoy-2011,c-zhang-2011, zhai-2011, hu-2011} 
or for interacting bosons~\cite{santos-2011, baym-2012}. 
Unfortunately, the experimental study of Rashba SOC requires 
more lasers and further developments are necessary to overcome several 
difficulties~\cite{dalibard-2011}. Thus, presently, artificial Rashba
SOC has not yet been created in the context of ultra-cold atoms. 
However, simultaneous theoretical studies of superfluidity for the 
experimentally relevant ERD spin-orbit coupling were performed 
for ultra-cold bosons by others~\cite{ho-2011, stringari-2012} 
and for ultra-cold fermions by our group~\cite{li-2012, seo-2012a, seo-2012b}.

One of the benchmarks of experimental studies of Fermi superfluidity of cold
atoms without artificial spin-orbit coupling was the emergence of molecular 
bound states via the use of Feshbach resonances~\cite{grimm-2010}, 
which lead to the formation of molecules~\cite{jin-2003a} and their posterior 
Bose-Einstein condensation in $^{40}{\rm K}_2$~\cite{jin-2003b} 
and $^6{\rm Li}_2$~\cite{grimm-2003}. In the present published literature
of ultra-cold fermions with spin-orbit coupling, only non-interacting
systems have been investigated~\cite{chinese-2012, zwierlein-2012}. 
However, very recently, the NIST group~\cite{spielman-2013} 
has demonstrated experimentally the formation of Feshbach molecules 
of ultra-cold fermions ($^{40}$K) in the presence of artificial SOC.
Our theoretical  results, described next, are in excellent agreement 
with the NIST experiment.

To address the important issue of the emergence of Feshbach
molecules for interacting fermions in the presence of artificial
SOC and Zeeman fields, we start from the 
Hamiltonian for two non-interacting fermions
\begin{equation}
H_0 
= 
H_1 
+ 
H_2,
\end{equation}
written as the sum of two contributions, which have the generic form
(with $\hbar = 1$)
\begin{equation}
H_j 
= 
\frac{{\bf \hat k}_j^2}{2m} 
-  
\left[
\left(
{\bf h}_R + {\bf h}_D 
\right)_j 
\cdot 
{\bf \sigma}_j
\right] 
- 
{\bf h}
\cdot 
{\bf \sigma}_j.
\end{equation}
The term containing 
$
{\bf h}_R 
= 
v_R
\left(
{\hat k}_x{\bf e}_y 
- 
{\hat k}_y {\bf e}_x 
\right)
$
represents the Rashba spin-orbit field, 
the term containing
$
{\bf h}_D 
= 
v_D 
\left(
{\hat k}_x {\bf e}_y 
+ 
{\hat k}_y {\bf e}_x 
\right)
$
represents the Dresselhaus spin-orbit field 
and
$
{\bf h}
= 
h_y {\bf e}_y
+
h_z {\bf e}_z
$
is the Zeeman field with $h_y = -\delta/2$
representing the detuning $\delta$ and $h_z = -\Omega_R/2$ 
representing the Raman intensity $\Omega_R$.
All these fields are described in energy units.
In addition, 
$
{\bf \hat k}_j 
= 
-i \nabla_j
$
is the momentum operator of the j-th particle, and 
$
{\bf \sigma}_j 
= 
\sigma_{x,j}{\bf e}_x
+ 
\sigma_{y,j}{\bf e}_y  
+
\sigma_{z,j}{\bf e}_z
$
is the vector Pauli matrix. 

\begin{figure} [tbh]
\centering
\epsfig{file=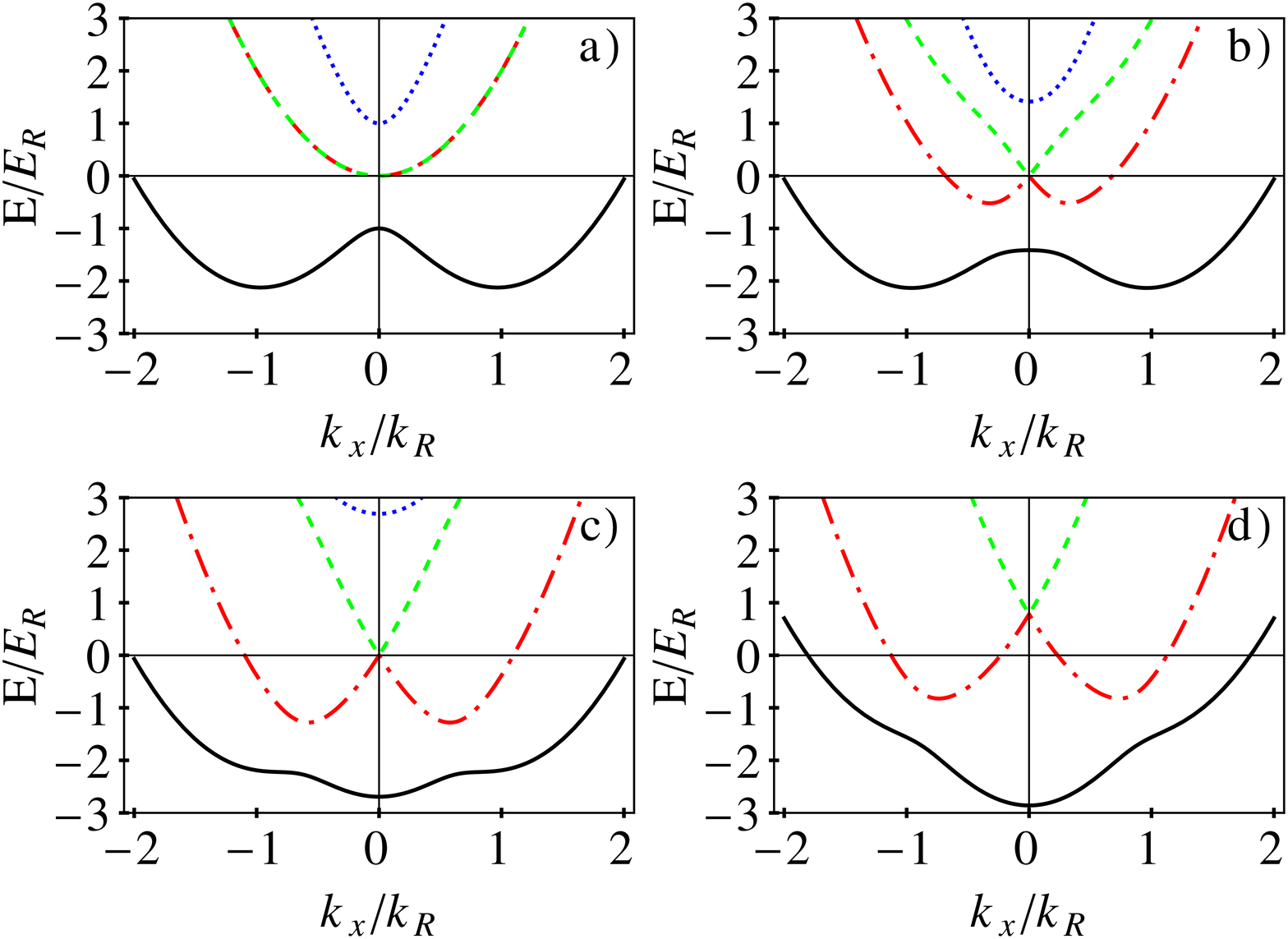,width=1.0 \linewidth}
\caption{ \label{fig:one}
(color online) 
Plots of the generalized two-particle helicity bands 
$E_{\Uparrow\Uparrow} ({\bf k}, {\bf K})$ (black solid),
$E_{\Uparrow\Downarrow} ({\bf k}, {\bf K})$ (red dot-dashed),
$E_{\Downarrow\Uparrow} ({\bf k}, {\bf K})$ (green dashed), 
and
$E_{\Downarrow\Downarrow} ({\bf k}, {\bf K})$ (blue dotted)
along the direction of relative momentum $(0, 0, k_x)$, 
for ERD spin-orbit coupling $v = k_R/m$,
various values of detuning $h_y$ and Raman intensity $h_z$, 
and specific values of the center of mass momentum $(0,0, K_x)$.
The parameters used are 
a) $h_y = 0$, $h_z = 0.5 E_R$, and $K_x = 0$; 
b) $h_y = 0.5 E_R$, $h_z = 1.0 E_R$, and $K_x = 0$;
c) $h_y = 1.25 E_R$, $h_z = 0.5 E_R$, and $K_x = 0$; 
and
d) $h_y = 0.5 E_R$, $h_z = 0.5 E_R$, and $K_x = 1.25 k_R$.
Notice the change in location of the minimum of $E_{\Uparrow\Uparrow}$ 
from finite $k_x$ in a) and b) to $k_x = 0$ in c) and d).
}
\end{figure}

The interaction Hamiltonian 
$
H_I ( {\bf r}_1, {\bf r}_2 )
=
- g \delta ( {\bf r}_1 - {\bf r}_2 )
\delta_{s_1 + s_2, 0}
$
describes zero-ranged attractive s-wave interactions between 
fermions with opposite spins $s_1 = -s_2$.
The bare coupling constant $g$ is renormalized through the use
of the Lippman-Schwinger relation 
$
L^3/g = -m/4\pi a_s + \sum_k 1/(2 \epsilon_k),
$ 
where $L^3$ is the volume,  
$
a_s 
= 
a_{bg}
\left[ 
1 + \Delta B/(B - B_0) 
\right]
$
is the three-dimensional scattering length expressed in terms
of the background scattering length $a_{bg}$, the real magnetic
field $B$, the width $\Delta B$ of the Feshbach resonance,
and the field $B_0$, where the resonance is located. 
While in the two-dimensional case, the bare coupling constant $g$
is eliminated in favor of the bare binding energy $E_{bs}$ via
$
L^2/g 
= 
\sum_{\bf k} 1/(2\epsilon_{\bf k} - E_{bs}).
$

Writing our Hamiltonian in relative momentum 
$
{\bf k} 
= 
({\bf k_1} - {\bf k_2})/2,
$
and center of mass (CM) momentum 
$
{\bf K} 
= 
{\bf k_1} + {\bf k_2}
$
coordinates and performing a global SU(2) spin rotation into 
singlet and triplet channels of the initial spin basis leads
to the non-interacting Hamiltonian matrix
\begin{eqnarray} 
\label{eqn:hamiltonian-singlet-triplet-basis}
{\bf H}_0 
= 
\left(
\begin{array}{c c c c}
\epsilon_{12} - 2h_z & -h_{12s}^*      &  0                    & -h_{12d}^*    \\
 -h_{12s}            &  \epsilon_{12}  & -h_{12s}^*            &  0            \\
 0                   & -h_{12s}        &  \epsilon_{12} + 2h_z & -h_{21d}      \\
-h_{12d}             &  0              & - h_{21d}^*           & \epsilon_{12}
\end{array} 
\right), 
\nonumber
\end{eqnarray}
where
$
\epsilon_{12} 
= 
\epsilon_{12} ({\bf k}, {\bf K})
=
\epsilon_{{\bf k}_1} + \epsilon_{{\bf k}_2}
$
is the sum of the kinetic energy of the two particles,
$
h_{12s} 
=
h_{12s} ({\bf k},{\bf K})
=
\left[
( h_{x1} + h_{x2} )
+ 
i (h_{y1} + h_{y2})
\right]
/\sqrt{2}
$
is the scaled sum of the tranverse fields felt by both particles,
$
h_{12d} =   
h_{12d} ({\bf k}, {\bf K})
=
\left[
( h_{x2} - h_{x1} )
+ 
i (h_{y2} - h_{y1})
\right]
/\sqrt{2}
$
and 
$
h_{21d} =   
h_{21d} ({\bf k}, {\bf K})
=
\left[
( h_{x1} - h_{x2} ) 
+ 
i (h_{y1} - h_{y2})
\right]
/\sqrt{2}
$
are scaled differences of the transverse fields felt by the
particles, where $h_{12d} = - h_{21d}$.
More explicitly
$
\epsilon_{12} ({\bf k},{\bf K}) 
=  
{k^2}/{m} + {K^2}/{(4m)};
$
the matrix element from total spin projection 
$m_s = 0 \to m_s = 1$ or $m_s = -1 \to m_s =  0$ in the triplet sector is
$
h_{12s} 
=
\left[
{\beta}K_y 
+
i 
(2 h_y + {\gamma}K_x )
\right]
/ 
\sqrt{2};
$
and the  matrix element that couples the singlet $(S = 0)$ and 
triplet $(S = 1)$ sectors with changes in total spin projection 
from $m_s = 0 \to m_s = 1$ is 
$
h_{12d} 
=
-\sqrt{2}\beta k_y 
- i \sqrt{2} \gamma k_x,
$
while the one with changes from $m_s = 0 \to m_s = -1$ 
is 
$
h_{21d}
=
\sqrt{2} \beta k_y
+ 
i \sqrt{2} \gamma k_x.
$
The variables $\gamma$ and $\beta$ are defined in terms of the Rashba $(v_R)$
and Dresselhaus $(v_D)$ coefficients as 
$\gamma = v_D + v_R$ and $\beta = v_D - v_R$. 
Notice that $H_0$ is not Galilean invariant, and that 
$h_{12s}$ depends only on the CM momentum ${\bf K}$,
while $h_{12d}$ and $h_{21d}$ depend only on the relative momentum ${\bf k}$, 
however the CM and relative coordinates remain coupled. 
Furthermore, in the experimentally relevant ERD 
case $v_D = v_R = v/2$ leading to $\beta = 0$ and $\gamma = v$.

Conditions for the emergence of Feshbach molecules
are obtained from the Schroedinger equation 
\begin{equation} 
\label{eqn:two-body-singlet-triplet-channel}
\left( 
{\bf H}_0 
+ 
{\bf H}_I 
\right)  
{\bf \Lambda}_{{\bf k},{\bf K}}  
 =  
E{\bf I}  
{\bf \Lambda}_{{\bf k}, {\bf K}}, 
\end{equation}
where the four-dimensional spinor 
$
{\bf \Lambda}_{{\bf k}, {\bf K}} 
= 
\left[
\Lambda_{1,1}({\bf k}, {\bf K}),
\Lambda_{1,0}({\bf k}, {\bf K}),
\Lambda_{1,-1}({\bf k}, {\bf K}),
\Lambda_{0,0}({\bf k}, {\bf K}) 
\right]
$
that includes both the triplet wavefunctions 
$
\Lambda_{1,1}({\bf k},{\bf K}) 
=
\psi_{\uparrow \uparrow}({\bf k},{\bf K})
$
corresponding to $(S = 1, m_s = 1)$;
$
\Lambda_{1,0}({\bf k},{\bf K}) 
=
\left[
\psi_{\uparrow \downarrow}({\bf k},{\bf K})
+
\psi_{\downarrow \uparrow}({\bf k},{\bf K}) 
\right] 
/\sqrt{2}
$
corresponding to $(S = 1, m_s = 0)$;
$
\Lambda_{1,-1}({\bf k},{\bf K}) 
= 
\psi_{\downarrow \downarrow}({\bf k},{\bf K})
$
corresponding to $(S = 1, m_s = -1)$;
and the singlet wavefunction
$
\Lambda_{0,0}({\bf k},{\bf K}) 
= 
\left[
\psi_{\uparrow \downarrow}({\bf k},{\bf K})
-\psi_{\downarrow \uparrow}({\bf k},{\bf K})  
\right]
/
\sqrt{2}
$
corresponding to $(S = 0, m_s = 0)$.
In this basis, the action of the interaction Hamiltonian on the 
four-dimensional spinor leads to the vector 
$
{\bf H}_I {\bf \Lambda}_{{\bf k},{\bf K}} 
=  
\left[ 
0,0,0,-g\Sigma_{\bf k} \Lambda_{0,0}(\bf k, \bf K) 
\right].
$

The equation above can be formally solved in any dimension 
by rearranging the Hamiltonian as 
$
\left[
E{\bf I}
-
{\bf H}_0 
\right]
{\bf \Lambda}_{{\bf k}, {\bf K}}
= 
{\bf H}_I {\bf \Lambda}_{{\bf k}, {\bf K}}
$
to obtain the relation 
\begin{equation}
{\bf \Lambda}_{{\bf k}, {\bf K}}
=
\frac{{\rm Adj}[E{\bf I} - {\bf H}_0]}
{{\rm Det} [E{\bf I} - {\bf H}_0]}
{\bf H}_I {\bf \Lambda}_{{\bf k}, {\bf K}},
\end{equation}
where ${\rm Adj}[{\bf M}]$ is the adjucate matrix 
and ${\rm Det}[{\bf M}]$ is the determinant of ${\bf M}$. 
Integration over the relative momentum ${\bf k}$ 
leads to the integral equation 
\begin{equation} 
\label{eqn:self-consistent-bound-state-energy}
\frac{L^d}{g} 
= 
-
\sum \limits_{\bf k} 
\frac
{
(
E - \epsilon_{12}
)
(
E - \epsilon_{12} + \vert h_t \vert
)
(
E - \epsilon_{12} - \vert h_t \vert
)
}
{
(E - E_1)(E - E_2)(E - E_3)(E - E_4) 
}.
\end{equation}

Here, the function 
$
\vert h_t \vert
=
\sqrt{
4h_z^2
+
\vert 
h_{12s}({\bf k}, {\bf K})
\vert^2
}
$
is the amplitude of the total field 
${\bf h}_t = {\bf h}_1 + {\bf h}_2$,
$d$ is the dimension of the system, and $E_i({\bf k}, {\bf K})$ are
the eigenvalues of ${\bf H}_0$, corresponding to the two-particle generalized
helicity bands $E_{\alpha \beta} ({\bf k}, {\bf K})$. 
We identify the right-hand-side 
of Eq.~(\ref{eqn:self-consistent-bound-state-energy}) with
the function 
$
G_s (E, {\bf K}) 
= 
\sum_{{\bf k},\alpha, \beta} 
\vert {\bf U}_{\alpha\beta,s}({\bf k},{\bf K})\vert^2
/
(E - E_{\alpha\beta} ({\bf k}, {\bf K})),
$
which corresponds to the spectral representation 
of the two-body Green's function for non-interacting fermions in the
singlet channel of the original spin states $(\uparrow, \downarrow)$.
Here, 
$
\vert 
{\bf U}_{\alpha \beta,s} ({\bf k}, {\bf K})
\vert^2
$ 
represents the spectral weight in the singlet channel $(s)$
associated with the spinor eigenvector 
${\bf U}_{\alpha \beta} ({\bf k}, {\bf K})$ of 
${\bf H}_0$ with eigenvalue 
$
E_{\alpha \beta} ({\bf k}, {\bf K}).
$
Only, the singlet channel contributes to $G_s ({\bf k}, {\bf K})$, 
as the interactions between fermions are non-zero only between 
the original $\uparrow$ and $\downarrow$ spins.

By ordering the eigenvalues 
$E_1 \ge E_2 \ge E_3 \ge E_4$, a simple inspection of 
Eq.~(\ref{eqn:self-consistent-bound-state-energy}) shows that 
a necessary condition for the formation of singlet Feshbach molecules 
occurs when
$
E({\bf K}) 
\le 
{\rm min}_{\bf k} 
\{
E_4 ({\bf k}, {\bf K})
\},
$
provided that there is spectral weight in the singlet interaction channel
for the lowest energy of two free fermions. 
The energies
$
E_{\alpha\beta} ({\bf k}, {\bf K})
$
can be written in terms of the two-particle kinetic energy 
$
\epsilon_{12} ({\bf k}, {\bf K}) 
$
and the effective fields
$
h_s ({\bf k}, {\bf K})
=
\vert h ({\bf k}_1) \vert
+
\vert h ({\bf k}_2) \vert,
$
and 
$
h_d ({\bf k}, {\bf K})
=
\vert h ({\bf k}_1) \vert
-
\vert h ({\bf k}_2) \vert,
$
where 
$
\vert 
h ( {\bf k}_i ) 
\vert
= 
\sqrt{
h_x^2 ( {\bf k}_i )
+
h_y^2 ( {\bf k}_i )
+
h_z^2 ( {\bf k}_i )
}
$
is the magnitude of the total field (spin-orbit and Zeeman) 
felt by the $i^{th}$ particle. The momenta are 
${\bf k}_1 = {\bf k} + {\bf K}/2$ and 
${\bf k}_2 = -{\bf k} + {\bf K}/2$. 
The eigenenergies for two free fermions are
$
E_{\Uparrow\Uparrow} ({\bf k}, {\bf K})
= 
\epsilon_{12} ({\bf k}, {\bf K}) 
- h_s ({\bf k}, {\bf K});
$
$
E_{\Uparrow\Downarrow} ({\bf k}, {\bf K})
= 
\epsilon_{12} ({\bf k}, {\bf K}) 
- 
h_d ({\bf k}, {\bf K});
$
$
E_{\Downarrow\Uparrow} ({\bf k}, {\bf K})
= 
\epsilon_{12} ({\bf k}, {\bf K}) 
+ 
h_d ({\bf k}, {\bf K})
$
and
$
E_{\Downarrow\Downarrow} ({\bf k}, {\bf K})
= 
\epsilon_{12} ({\bf k}, {\bf K}) 
+ 
h_s ({\bf k}, {\bf K}).
$

From Eq.~(\ref{eqn:self-consistent-bound-state-energy}),
we obtained Feshbach molecule energies $E = E_B ({\bf K})$
for an arbitrary mixture of Rashba and Dresselhaus terms at any 
value of ${\bf K}$ in 2D and 3D.
We also calculated the effective mass tensor
and the corresponding Bose-Einstein condensation 
temperature~\cite{kurkcuoglu-2013}.
However, we will show here results only for the ERD case,
because of its experimental relevance for ultra-cold 
fermions~\cite{spielman-2011}. Respectively, we use as 
units of energy and momentum, the  
recoil energy $E_R = k_R^2/(2m)$ and the recoil momentum 
$k_R = 2\pi/\lambda$, where $\lambda$ is the wavelength of the 
laser light used in the Raman beams~\cite{spielman-2011}. 
We parametrize the ERD coupling parameter as
$
v 
= 
\left[
1 - \cos(\theta) 
\right]
k_R 
/(2m)
$
for Raman beams that cross at an arbitrary angle $\theta$. 
Current experiments correspond to $\theta = \pi$ and $v = k_R/m$.

To gain insight in the ERD case, first we solve the 
problem for zero CM momentum ${\bf K} = {\bf 0}$ 
and zero detuning $h_y = 0$, but finite $h_z$.
In this case, the eigenvalues take the simple form 
$
E_{\Uparrow \Uparrow} ({\bf k}, {\bf 0}) 
= 
k^2/m 
+ 
2 \vert h_{\rm eff} ({\bf k})\vert 
$
for the highest energy,
$
E_{\Uparrow \Downarrow} ({\bf k}, {\bf 0}) 
= 
E_{\Downarrow \Uparrow} ({\bf k}, {\bf 0}) 
= 
k^2/m, 
$
for the intermediate energies,
and 
$
E_{\Downarrow \Downarrow} ({\bf k}, {\bf 0}) 
= 
k^2/m 
- 
2 \vert h_{\rm eff} ({\bf k})\vert 
$
for the lowest energy,
where the magnitude of the effective field is 
$
\vert h_{\rm eff} ({\bf k})\vert 
=
\sqrt{(vk_x)^2 + h_z^2}.
$
The condition for the emergence of zero CM 
momentum Feshbach molecules is then 
$
E ({\bf K} = 0)
\le 
{\rm min}_{\bf k} 
\{
E_{\Uparrow\Uparrow} ({\bf k},{\bf 0})
\}.
$
See Fig.~\ref{fig:one} for examples of $E_{\alpha \beta} ({\bf k}, {\bf K})$ 
in various regimes.

\begin{figure} [tbh]
\centering
\epsfig{file=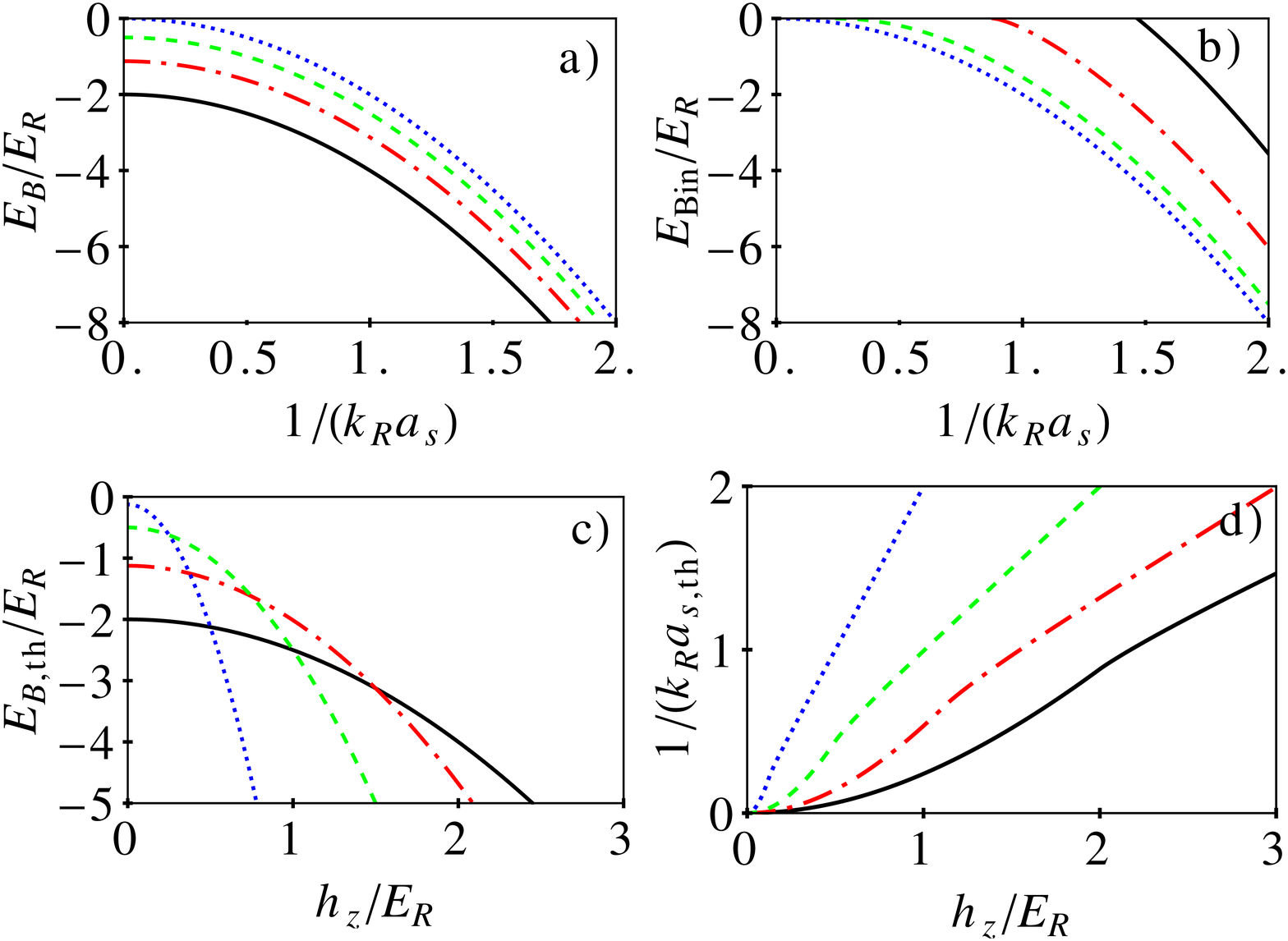,width=1.0 \linewidth}
\caption{
\label{fig:two}
(color online)
Plots of bound state energy $E_B/E_R$ versus
$1/(k_R a_s)$ with ${\bf K} = {\bf 0}$, 
$h_z = 0$, $h_y = 0$ 
are shown in a) 
for $v = 0$ (blue dotted), 
$v = 0.5 k_R/m$ (red dotdashed), 
$v = 0.75 k_R/m$ (green dashed),
and $v = k_R/m$ (black solid).
Plots of $E_{Bin}/E_R$ versus $1/(k_R a_s)$ with 
${\bf K} = {\bf 0}$, $v = k_R/m$, $h_y = 0$
are shown in b) 
for $h_z = 0$ (blue dotted),
$h_z = E_R$ (green dashed), 
$h_z = 2E_R$ (red dotdashed), 
$h_z = 3E_R$ (black solid).
Plots of $E_{B,th}/E_R$ and 
$1/(k_R a_{s,th})$ versus $h_z/E_R$ with ${\bf K} = {\bf 0}$, $h_y = 0$ 
are shown respectively in c) and d) 
for $v = 0.25 k_R/m$ (blue dotted)
$v = 0.5 k_R/m$ (red dotdashed)
$v = 0.75 k_R/m$ (green dashed)
$v =  k_R/m$ (black solid).
}
\end{figure}

In Fig.~\ref{fig:two}, we show the bound-state energy $E = E_B$ 
of Feshbach molecules (in units of $E_R$) versus the scattering parameter 
$1/(k_R a_s)$ at zero CM momentum $({\bf K} = 0)$ 
and zero detuning $(h_y = 0)$ in two situations.
In the limit of $h_z \to 0$, the Feshbach molecule energy can be 
obtained analytically as $E_B = -1/(ma_s^2) - mv^2$, which means that
the existence of spin-orbit coupling lowers the energy of the bound
state from the standard value $E_{B,0} = -1/(ma_s^2)$ 
by an amount equal to twice the kinetic energy transferred to 
individual atoms. However, the threshold scattering length for the emergence
of bound states remains at $a_s \to \infty$ or $1/(k_R a_s) = 0$.
The binding energy at ${\bf K = 0}$ is defined as
$
E_{Bin} 
=  
E_B 
- 
{\rm min}_{\bf k} E_{i} ({\bf k}, {\bf K} = {\bf 0})
$
is a better indicator of the effects of spin-orbit coupling
since the minimum energy of two free fermions also changes with $v$,
where $E_i \to E_{\Uparrow \Uparrow}$ when $h_z \ne 0$, but
$E_i \to E_{\Uparrow \Downarrow}$ when $h_z \to 0$ due to zero spectral
weight in $\Uparrow\Uparrow$ channel. An example of $E_{Bin}$ is shown
in Fig.~\ref{fig:two}b, where we plot $E_{Bin}/E_R$ versus $1/(k_R a_s)$ 
for fixed $v = k_R/m$ and increasing Zeeman field $h_z$.
As $h_z$ increases, the threshold for the formation 
of Feshbach molecules is shifted from infinite to finite 
and positive scattering lengths, indicating that stronger 
attraction between fermions is necessary to
overcome the depairing effects of $h_z$.

In Fig.~\ref{fig:two}c, we show the threshold bound-state energy $E_{B, th}$ 
for ${\bf K} = 0$ and $h_y = 0$ as a function of $h_z$ 
and varying $v$, obtained
from the threshold condition 
$
E ({\bf K}) 
\le 
{\rm min}_{\bf k} \
\{
E_{\Uparrow \Uparrow}({\bf k},{\bf K})
\}.
$
For $({\bf K} = 0)$,
$
E_B 
\le 
E_{B, th} 
= 
 mv^2
- 
h_z^2/(m v^2), 
$
having the dimensionless form
$
{\widetilde E}_{B, th}
=
2 {\widetilde p}^2 
- 
{\widetilde h}_z^2/(2 {\widetilde p}^2),
$
where ${\widetilde E}_{B, th} = E_{B,th}/E_R$, 
${\widetilde h}_z = h_z/E_R$, and 
${\widetilde p} = p/k_R$ with $p = mv$.
In Fig.~\ref{fig:two}d, we show the threshold scattering length $a_{s,th}$ 
as a function of $h_z$, which behaves differently 
as $h_z$ reaches the critical value $h_{z,c} = mv^2$. 
This condition is expressed in dimensionless units as 
${\widetilde h}_{z,c} = 2 {\widetilde p}^2$.
It is at this critical value that 
$
E_{\Uparrow\Uparrow} ({\bf k}, {\bf K})
$
changes from a double minimum
when ${\widetilde h_z} \le {\widetilde h}_{z,c}$ to a single minimum when
${\widetilde h_z} \ge {\widetilde h}_{z,c}$.
For fixed SOC $v$, the threshold $a_{s,th}$ 
progressively grows with increasing $h_z$ first quadratically 
for $h_z < h_{z,c}$ and then linearly for $h_z > h_{z,c}$,
as stronger attractive s-wave (singlet) interactions are necessary 
to overcome the de-pairing effect of $h_z$ that tends to align the 
original spins.

\begin{figure} [tbh]
\centering
\epsfig{file=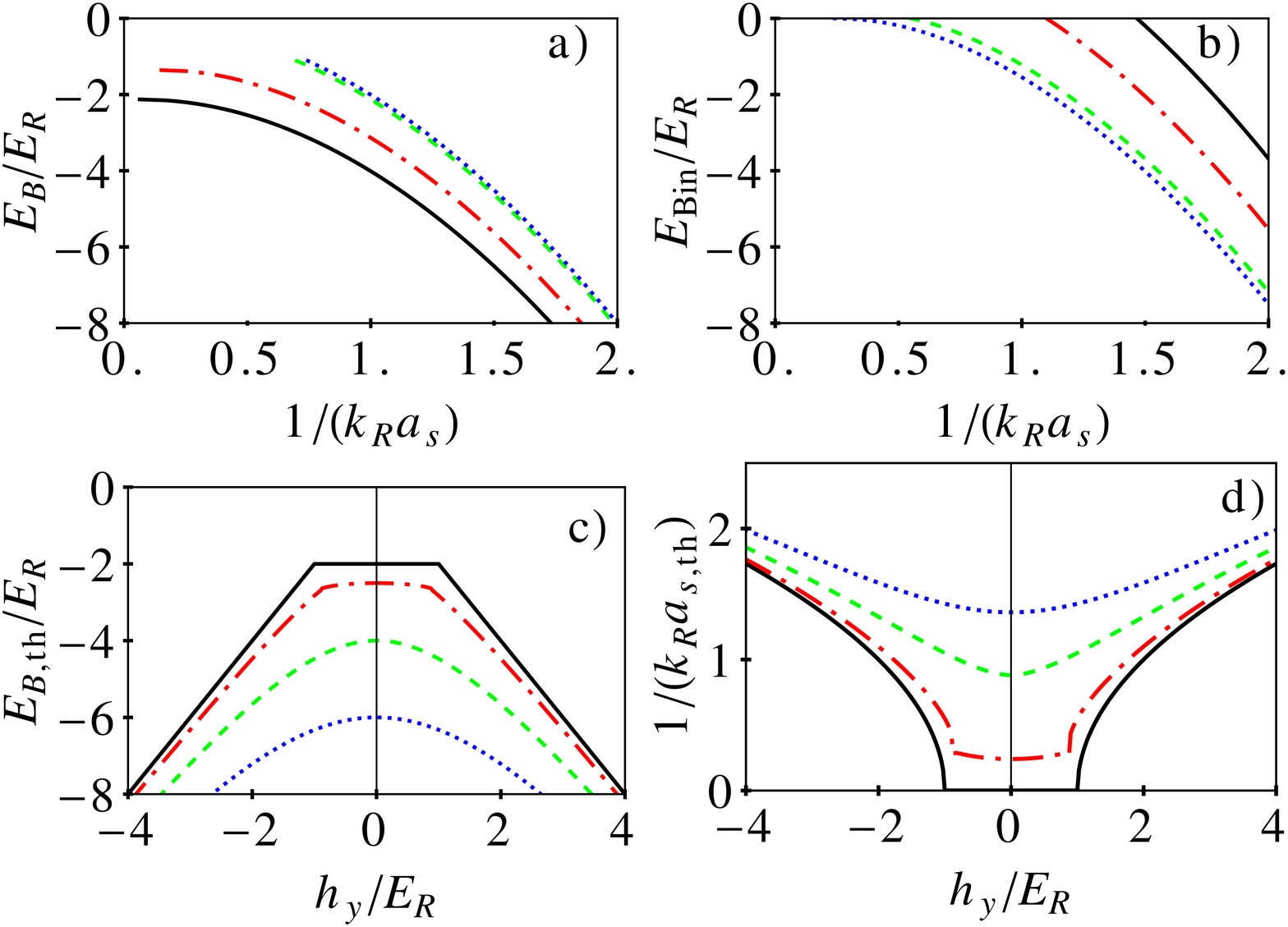,width=1.00 \linewidth}
\caption{
\label{fig:three}
(color online)
a) Plots of $E_B/E_R$ versus $1/(k_R a_s)$ 
with ${\bf K} = 0$, $h_y = 0.25E_R$, and $h_z = 0.5E_R$ for 
$v = 0.05 k_R/m$ (blue dotted),
$v = 0.25 k_R/m$ (red dotdashed),
$v = 0.75 k_R/m$ (green dashed),
$v = k_R/m$ (black solid).
b) Plots of $E_{Bin}/E_R$ versus $1/(k_R a_s)$ 
with ${\bf K} = 0$, $h_z = E_R$, and $v = k_R/m$ 
for $h_y = 0$ (blue dotted),
$h_y = E_R$ (green dashed),
$h_y = 2E_R$ (red dotdashed),
$h_y = 3E_R$ (black solid).
Plots of $E_{B, th}/E_R$ and 
$1/(k_R a_{s,th})$  versus $h_y/E_R$ with
${\bf K} = {\bf 0}$, and $v = k_R/m$ are shown 
respectively in c) and d) for 
$h_z = 0$ (black solid);
$h_z = E_R$ (red dotdashed);
$h_z = 2 E_R$ (green dashed);
$h_z = 3 E_R$ (blue dotted). 
}
\end{figure}

In Fig.~\ref{fig:three}, we show some results 
for finite $h_y$ and ERD spin-orbit coupling $v$, 
but zero CM momentum $( {\bf K} = 0 )$. 
Fig.~\ref{fig:three}a contains plots of the Feshbach molecule energy 
$E_B/E_R$ versus interaction parameter $1/(k_R a_s)$
for $h_y = 0.25E_R$ and $h_z = 0.5E_R$ and changing ERD spin-orbit 
coupling $v$. The threshold interaction parameter 
decreases with increasing $v$ showing that larger SOC facilitates
the formation of molecules when Zeeman fields are present.
In Fig.~\ref{fig:three}b, we show $E_{Bin}/E_R$ 
versus $1/(k_R a_s)$ for $v = k_R/m$, 
$h_z = E_R$, and ${\bf K} = 0$, but changing $h_y$. 
The scattering parameter threshold increases with $h_y$ 
as a stronger attractive interaction is necessary  
to form singlet Feshbach molecules in the original 
spin basis $(\uparrow, \downarrow)$. 
The bound state energy threshold $E_{B, th}/E_R$
and the scattering parameter threshold $1/(k_R a_{s,th})$ 
versus $h_y/E_R$ are shown respectively in Figs.~\ref{fig:three}c 
and~\ref{fig:three}d for ${\bf K} = {\bf 0}$, and $v = k_R/m$, but changing 
$h_z$. 

\begin{figure} [tbh]
\centering
\epsfig{file=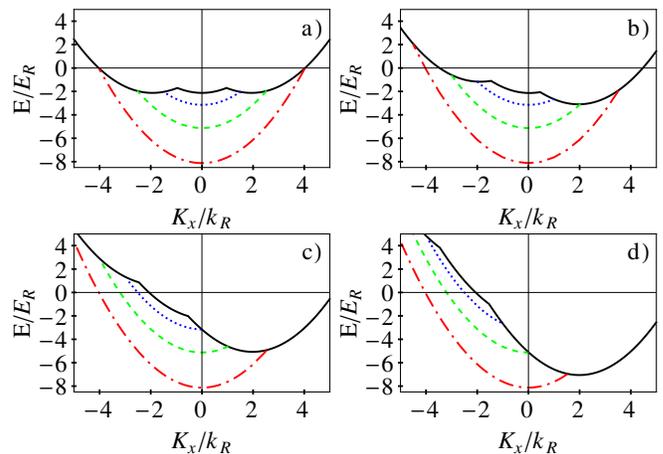,width=1.0 \linewidth}
\caption{ \label{fig:four}
(color online)
Plots of bound state threshold energies (solid black) 
and of energies of Feshbach molecules 
(blue dotted with $1/(k_R a_s) = 0.75$; 
green dashed with $1/(k_R a_s) = 1.25$;
red dotdashed with $1/(k_R a_s) = 1.75$) 
versus center of mass momentum ${\bf K} = (0, 0, K_x)$ 
for $v = k_R/m$ and $h_z = 0.5E_R$, with  
a) $h_y = 0$;
b) $h_y = 0.5 E_R$; 
c) $h_y = 1.5E_R$; 
and 
c) $h_y = 2.5E_R$.
Notice the absence of inversion symmetry (parity) when $h_y \ne 0$.}
\end{figure}

In Fig.~\ref{fig:four}, we show the energy dispersions $E({\bf K})$ of Feshbach
molecules and the threshold energy
$
E_{B,th} 
=
{\rm min}_{\bf k}
\{
E_{\Uparrow\Uparrow} ({\bf k}, {\bf K})
\}
$
along CM momentum ${\bf K} = (K_x, 0, 0)$ for four different
cases with fixed scattering parameter $1/(k_R a_s) = 0.75$,
spin-orbit coupling $v = k_R/m$, and Raman intensity $h_z = 0.5 E_R$.
In Fig.~\ref{fig:four}a,  $h_y = 0$, while in b, c and d, the
values are $h_y = 0.5E_R, 1.5E_R$, and $2.5E_R$, respectively.
Only in Fig.~\ref{fig:four}a the dipersions are even in $K_x$ since $h_y = 0$,
otherwise, for $h_y \ne 0$, parity is lost. In particular, Feshbach molecules
are stable only in a range of CM momenta, outside this region they decay into
the two-particle continuum due to Landau damping.
It is important to notice that for $h_y$ positive and increasing,
the region of stable Feshbach molecules shifts towards negative
CM momenta, and that beyond a critical value $h_{y,c}$ no Feshbach molecules 
with zero CM momentum are stable. This unusual effect is a direct consequence 
of the absence of Galilean invariance and the loss of parity.

We have investigated the emergence of Feshbach molecules
in the presence of spin-orbit coupling and Zeeman fields for any mixture
of Rashba and Dresselhaus terms, but focused on the experimentally relevant
equal Rashba and Dresselhaus (ERD) spin-orbit coupling. For
zero detuning $(h_y = 0)$ and fixed ERD spin-orbit coupling, 
we have found that the threshold scattering parameter $[1/(k_F a_{s,th})]$ 
required to form Feshbach molecules with zero center-of-mass (CM) momentum 
is shifted to larger positive values
when the Raman coupling $(h_z)$ is increased. 
Furthermore, for fixed scattering parameter $[1/(k_F a_s)]$ 
these molecules are stable only for a symmetric range of CM momenta, 
outside which they decay into the two-particle continuum. 
However, for finite detuning $(h_y \ne 0)$ and fixed 
Raman intensity, ERD spin-orbit coupling and 
scattering parameter, Feshbach molecules are stable only 
in an asymmetric range of CM momenta, and if the detuning 
is sufficiently large Feshbach molecules with zero CM momentum are 
not possible.  These effects are a manifestation of the absence of 
Galilean invariance and the loss of parity. 

\acknowledgments{We thank ARO (W911NF-09-1-0220) for support, 
Ian Spielman and Ross Williams for discussions,
and Vijay Shenoy for alerting us to his related work~\cite{shenoy-2013}.}


\begin{thebibliography}{2}

%
%%
\bibitem{kane-2010}
M. Z. Hasan and C. L. Kane,
Rev. Mod. Phys. {\bf 82}, 3045 (2010).

%
%%
\bibitem{s-zhang-2011}
X. L. Qi and S. C. Zhang,
Rev. Mod. Phys. {\bf 83}, 1057 (2011).

%
%%
\bibitem{spielman-2011} 
Y. J. Lin, K. Jimenez-Garcia, and I. B. Spielman, 
Nature {\bf 471}, 83 (2011).

%
%%
\bibitem{chapman-2011} 
M. Chapman, and C. S\'a de Melo, 
Nature {\bf 471}, 41 (2011).

%
%%
\bibitem{shenoy-2011} 
J. P. Vyasanakere, S. Zhang, and V. B. Shenoy, 
Phys. Rev. B {\bf 84}, 014512 (2011).

%
%%
\bibitem{c-zhang-2011}
M. Gong, S. Tewari, and C. Zhang, 
Phys. Rev. Lett. {\bf 107}, 195303 (2011).

%
%%
\bibitem{zhai-2011}
Z.-Q. Yu and H. Zhai, 
Phys. Rev. Lett. {\bf 107}, 195305 (2011).

%
%%
\bibitem{hu-2011}
H. Hu, L. Jiang, X.-J. Liu, and H. Pu, 
Phys. Rev. Lett. {\bf 107}, 195304 (2011). 


%
%%
\bibitem{santos-2011} 
S. Sinha, R. Nath, and L. Santos, 
Phys. Rev. Lett. {\bf 107}, 270401 (2011).

%
%%
\bibitem{baym-2012} 
T. Ozawa, and G. Baym,
Phys. Rev. Lett. {\bf 110}, 085304 (2013). 
%
%% Condensation transition of ultracold Bose gases with Rashba 
%% spin-orbit coupling
%

%
%%
\bibitem{dalibard-2011}
 J. Dalibard, F. Gerbier, G. Juzeliunas, and P. Ohberg, 
Rev. Mod. Phys. {\bf 83}, 1523 (2011).

%
%%
\bibitem{ho-2011}
T. L. Ho and S. Zhang,
Phys. Rev. Lett. {\bf 107}, 150403 (2011).
%
%%Bose-Einstein Condensates with Spin-Orbit Interaction
%

%
%%
\bibitem{stringari-2012}
Y. Li, L. P. Pitaevskii, S. Stringari,
Phys. Rev. Lett. {\bf 108}, 225301 (2012). 
%
%% Quantum tri-criticality and phase transitions in spin-orbit 
%% coupled Bose-Einstein condensates
%

%
%%
\bibitem{li-2012}
L. Han, and C. A. R. S\'a de Melo,
Physical Review A {\bf 85}, 011606 (R) (2012).

%
%%
\bibitem{seo-2012a}
K. Seo, L. Han, and C. A. R. S\'a de Melo,
Phys. Rev. A {\bf 85}, 033601 (2012). 

%
%%
\bibitem{seo-2012b}
K. Seo, L. Han, and C. A. R. S\'a de Melo,
Phys. Rev. Lett. {\bf 109}, 105303 (2012).

%
%%
\bibitem{grimm-2010}
C. Chin, R. Grimm, P. Julienne, and E. Tiesinga,
Rev. Mod. Phys. {\bf 82}, 1225 (2010). 
%
%% Feshbach resonances in ultracold gases
%

%
%%
\bibitem{jin-2003a}
C. A. Regal, C. Ticknor, J. L. Bohn, and D. S. Jin, 
Nature {\bf 424}, 47 (2003). 
%
%% Creation of ultracold molecules from a Fermi gas of atoms
%

%
%%
\bibitem{jin-2003b}
M. Greiner, C. A. Regal, and D. S. Jin, 
Nature {\bf 426}, 537 (2003).
%
%% Emergence of a molecular Bose-Einstein condensate from a Fermi gas
%

%
%%
\bibitem{grimm-2003}
S. Jochim, M. Bartenstein, A. Altmeyer, G. Hendl, S. Riedl, 
C. Chin, J. Hecker Denschlag, and R. Grimm
Science {\bf 302}, 2101 (2003).
%
%% Bose-Einstein condensation of molecules
%

%
%%
\bibitem{chinese-2012}
P. Wang, Z.-Q. Yu, Z. Fu, J. Miao, L. Huang, S. Chai,
H. Zhai, and J. Zhang, 
Phys. Rev. Lett. {\bf 109}, 095301 (2012).

%
%%
\bibitem{zwierlein-2012}
L. W. Cheuk, A. T. Sommer, Z. Hadzibabic, T. Yefsah,
W. S. Bakr, and M. W. Zwierlein, 
Phys. Rev. Lett. {\bf 109}, 095302 (2012).

%
%%
\bibitem{spielman-2013}
R. A. Williams, M. C. Beeler, L. J. LeBlanc, K. Jim\'enez-Garc\'ia 
and I. B. Spielman, to appear in the arXiv (2013).

%
%%
\bibitem{kurkcuoglu-2013}
D. M. Kurkcuoglu and C. A. R. S\'a de Melo, 
unpublished (2013).

%
%%
\bibitem{shenoy-2013}
V. B. Shenoy, arXiv:1211.1831.


\end{thebibliography}
\end{document}